\documentclass[aps,preprint,preprintnumbers,showpacs,groupedaddress,superscriptaddress,floatfix]{revtex4-1} 
\usepackage{amssymb}
\usepackage{amsmath}
\usepackage{graphicx}
\usepackage{geometry}
\begin{document}
\preprint{Contribution to EIC Users Group Meeting, Catholic University, Washington DC, July 2018}

\title{\Large \bf Spin-Dependent DIS from \\ Polarized $^3$He at EIC}
\vskip 0.25 true in
\author{\it Richard Milner \footnote{Email: milner@mit.edu}}
\affiliation{Physics Department and Laboratory for Nuclear Science, Massachusetts Institute of Technology, Cambridge, MA 02139}

\begin{abstract}
A day-1 EIC detector should be prepared to carry out inclusive DIS measurements from electron collisions with nucleon and nuclear beams.  Here I consider the scientific motivation for spin-dependent DIS measurements from a polarized $^3$He beam.  Longitudinally polarized $^3$He ions can be used to measure $g^n_1(x,Q^2)$ in purely inclusive scattering. I also consider spin-dependent DIS from polarized $^3$He by tagging the spectator proton or deuteron in coincidence with the scattered electron.   A calculation of high energy scattering from the polarized $^3$He nucleus in the EIC frame is motivated. Further, this type of measurement can be extended to other polarized few-body nuclei.  High energy spin-dependent scattering from polarized $^3$He with tagged proton and deuteron could be studied experimentally ahead of EIC using spin-dependent proton-$^3$He scattering in RHIC, when polarized $^3$He ions become available. 
\end{abstract}
\date{\today}
\maketitle

\section{Introduction}

\begin{figure}[!htb]
\centering\includegraphics[width=0.5\textwidth,height=0.25\textheight]{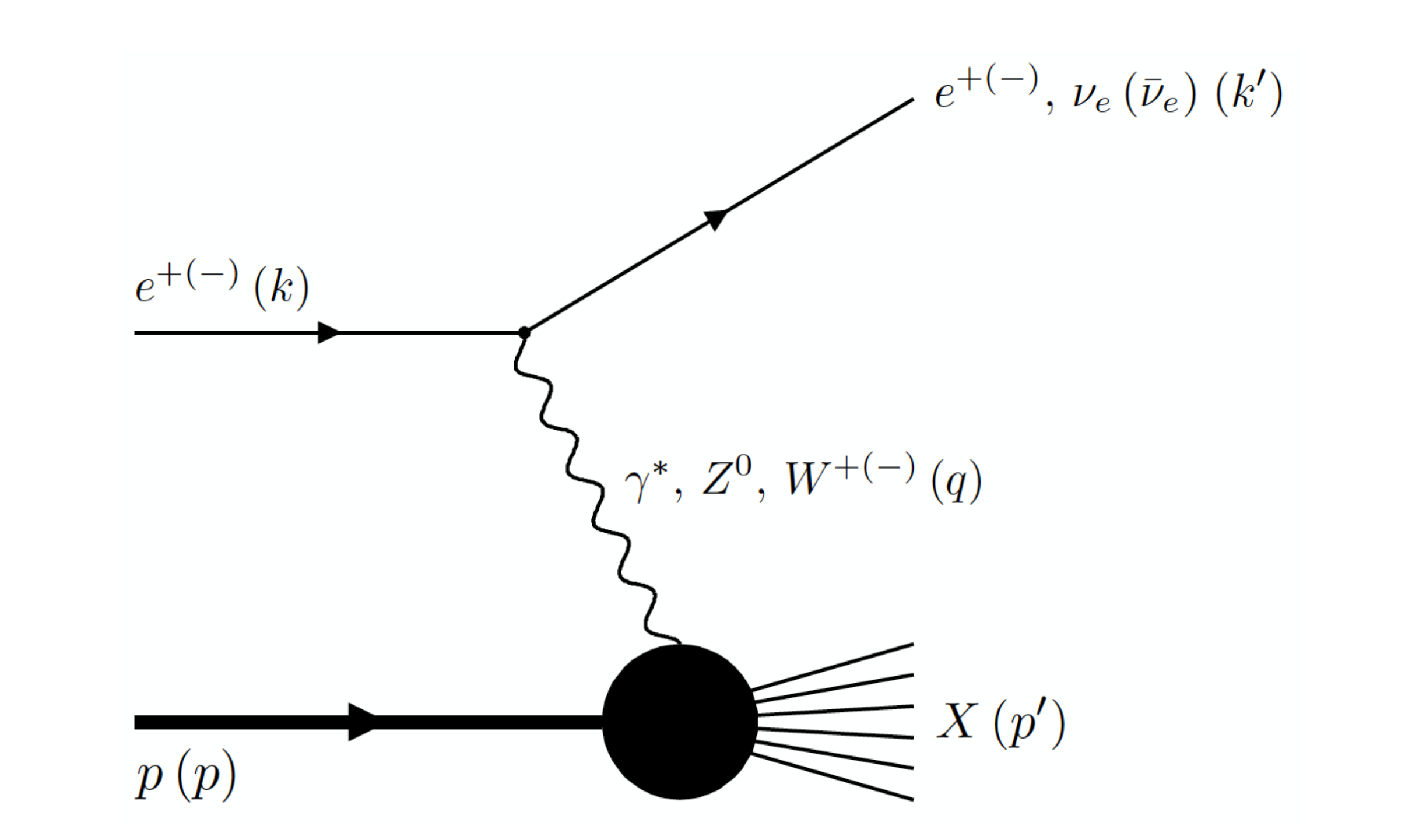}
\caption{Feynman diagram describing unpolarized DIS to lowest order perturbation theory.}
\label{dis}
\end{figure}

Following~\cite{Don2017}, the leading order Feynman diagram for electron$-$nucleon (mass $M$) scattering is shown in Fig.~\ref{dis}
with the electron (positron) and nucleon in the initial state denoted by the four-vectors
$k = (E_e; {\bf k_e})$ and $p = (E_P ;{\bf P})$, respectively. The  final state consists of the
scattered lepton $k^\prime = (E^\prime_l ; {\bf k^\prime_l})$ and the hadronic  final state system $p^\prime = (E_X; {\bf p_X})$.

Lepton-nucleon scattering is characterized in terms of a relativistic-invariant formulation summarized in the following variables:
\begin{eqnarray}
s &=& (k+p)^2 \approx 4 E_e E_N \\
t &=& (p-p^\prime)^2 \\
u &=& (k^\prime-p)^2 \\
Q^2 &=& -(k-k^\prime)^2 = -(p-p^\prime)^2 = -t  \\
x &=& \frac{Q^2}{2(p \cdot q)} \approx -\frac{t}{u+s} \ , \ \ \ 0 \le x \le 1 \\
y &=& \frac{p \cdot q}{p \cdot k} \approx \frac{u+s}{s} \ , \ \ \ 0 \le y \le 1 \\
W^2 &=& (p+q)^2 = (p^\prime)^2 = M^2 +\frac{Q^2}{x}(1-x) \approx s+t+u \\
\nu &=& \frac{p \cdot q}{M} \\ 
\end{eqnarray}
The $\approx$ sign refers to those cases where the electron and nucleon masses have been neglected.
Note that the center-of-mass energy for laboratory energies of $E_e$ (electron) and $E_N$ (nucleon) is 
$\sqrt{s} = 2 \sqrt{E_e E_N}$.  
$Q^2$ is the negative square of the 4-momentum transfer $q$ and denotes the virtuality of the exchanged gauge boson. The momentum transfer $q$ determines the size of the wavelength of the virtual boson and therefore
the object size which can be resolved in the scattering process. To resolve objects
of size $\Delta$ requires the wavelength of the virtual boson $\lambda$ to be smaller than $\Delta$.
$W^2$ is the square of the invariant mass of the hadronic final state system $X$. $W$ can also be interpreted as the center-of-mass energy of the gauge boson-nucleon system. Small values of $x$ correspond to large values of the
invariant mass $W$. Recall, $x$ is the Bjorken scaling variable and is interpreted in
the quark-parton model as the fraction of the nucleon momentum carried by the
struck parton. The limits on $x$ follow from the fact that the square of the invariant
mass $W^2$ has to be larger or equal to the square of the mass of the nucleon $M^2$
i.e., $W^2 = M^2 + (Q^2/x)(1 - x) \ge M^2$ where $x$ = 1 corresponds to the elastic case for which $W=M$.

In the proton rest frame, $\nu$ is the energy of the exchanged photon.  The maximum energy transfer $\nu_{max}$ is given
by $\nu_{max} = s/2M$.  The quantity $y$ is the fraction of the incoming electron energy carried by the exchanged photon,
also known as the {\it inelasticity} in the rest frame of the nucleon.  $y$ can also be written as $y = \nu/\nu_{max}$.   The relativistic invariant variables $x, y, Q^2$ and $s$ (the center-of-mass energy squared) are connected via
$$
Q^2 \approx s \cdot x \cdot y \ ,
$$
where the electron and nucleon masses have been ignored.  For fixed $x$ and $y$, an $eN$
collider allows one to reach much larger values of $Q^2$ as well as much lower values
of $x$, keeping $y$ and $Q^2$ fixed, due to the larger center-of-mass energy $\sqrt s$ compared
with fixed-target experiments.  The minimum value of $x$ attainable for $Q^2 \ge$ 1 GeV$^2$ is
given by
$$
x_{min} = \frac{1}{s_{max} \cdot y_{max}} \ .
$$
Deep inelastic lepton-nucleon scattering is defined to occur, i.e. Bjorken scaling is observed experimentally,  when $Q^2 \ge$ 1 GeV$^2$ and $W >$ 2 GeV, to avoid the resonance region.  

When carried out with longitudinally polarized nucleons, DIS probes the helicity parton distribution
functions of the nucleon. For each flavor $f = u, d, s, \bar{u}, \bar{d}, \bar{s}, g$ these are defined by
$$
\Delta f(x,Q^2) \equiv  f^+(x,Q^2) - f^-(x,Q^2) \ , 
$$
with $f^+ (f^-)$ denoting the number density of partons with the same (opposite) helicity as the nucleons, as a function
of the momentum fraction $x$ and the resolution scale $Q$.   Similar to the unpolarized quark and gluon densities, the
$Q^2$-dependences of $\Delta q(x,Q^2), \Delta {\bar q}(x,Q^2)$ and the gluon helicity distribution $\Delta g(x,Q^2)$ are related by QCD radiative processes that are calculable. When integrated over all momentum fractions and appropriately
summed over flavors, the $\Delta f$ distributions give the quark and gluon spin contributions to the proton
spin which appear in the fundamental proton helicity sum rule.
\begin{figure}[!htb]
\centering\includegraphics[width=0.8\textwidth,height=0.8\textheight]{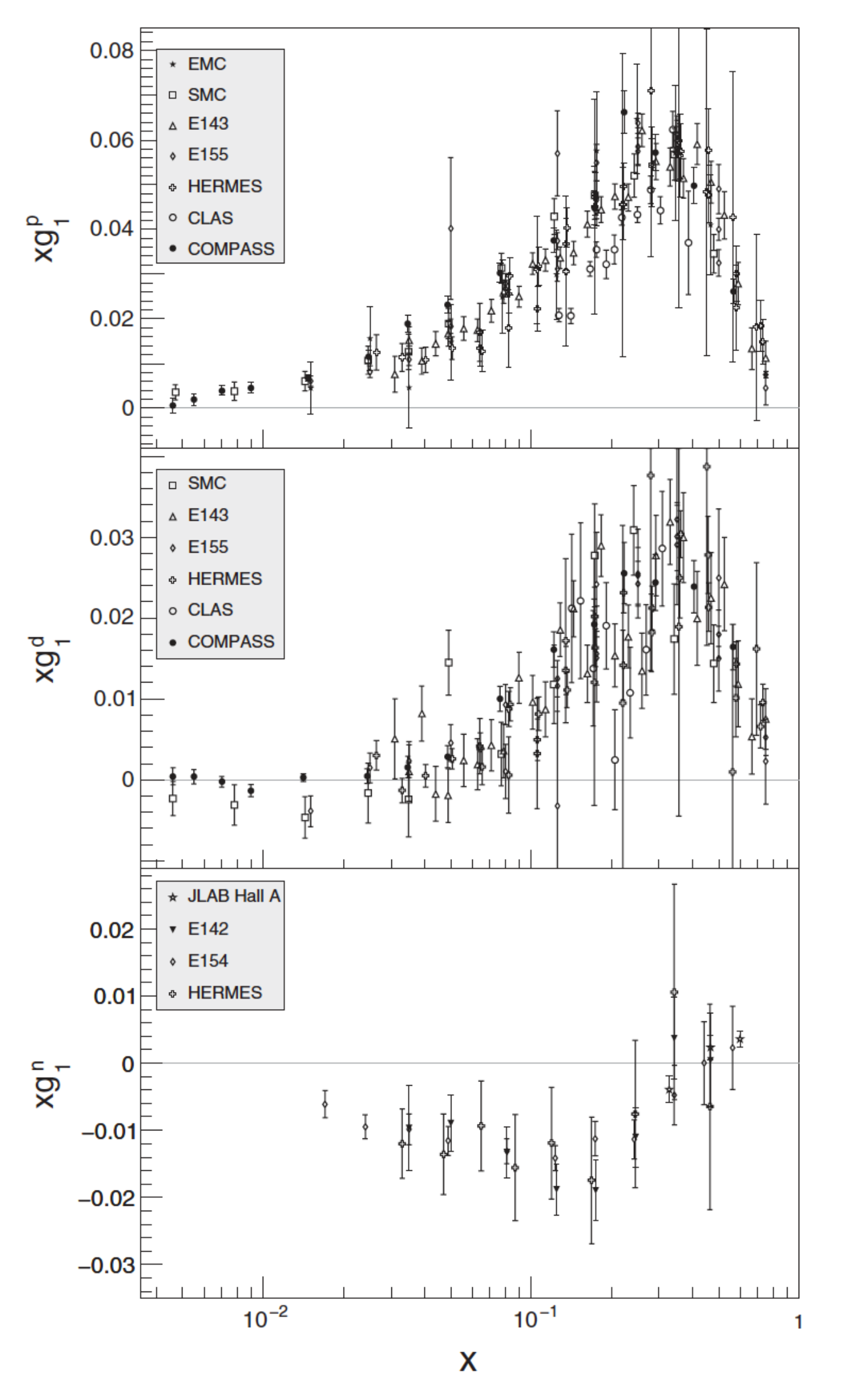}
\caption{From~\cite{Aid2013}, world data on $xg_1$ as a function of $x$ for the proton (top),
the deuteron (middle), and the neutron, i.e. polarized $^3$He, (bottom) at the $Q^2$ of the
measurement. Only data points for $Q^2 >$ 1 GeV $^2$ and $W > $ 2.5 GeV
are shown. Error bars are statistical errors only.}
\label{world}
\end{figure} 
Experimental access to the $\Delta f$ in lepton-scattering is obtained through the spin-dependent
structure function $g_1(x,Q^2)$, which appears in the polarization difference of cross-sections when the lepton
and the nucleon collide with their spins anti-aligned
or aligned:
$$
\frac{1}{2} \left[ \frac{d^2 \sigma^{\uparrow \downarrow}}{dx d Q^2} - \frac{d^2 \sigma^{\uparrow \uparrow}}{dx dQ^2}\right] \approx \frac{4 \pi \alpha^2}{Q^4} y(2-y) g_1(x,Q^2) \ .
$$
The expression above assumes photon exchange between
the lepton and the nucleon.   In leading order in the strong coupling constant, the
structure function $g_1(x,Q^2)$ of the proton can be written
as 
$$
g_1(x,Q^2) = \frac{1}{2} \sum e^2_q \left[ \Delta q(x,Q^2) + \Delta {\bar q}(x,Q^2) \right] \  , 
$$
where $e_q$ denotes a quark's electric charge. Higher order
expansions contain calculable QCD coefficient functions. The structure function $g_1(x,Q^2)$ is thus directly
sensitive to the nucleon spin structure in terms of
the combined quark and anti-quark spin degrees of freedom.   We define the scattering asymmetry 
$$
A_1 \equiv \frac{\left[ \frac{d^2 \sigma^{\uparrow \downarrow}}{dx d Q^2} - \frac{d^2 \sigma^{\uparrow \uparrow}}{dx dQ^2}\right]} 
{\left[ \frac{d^2 \sigma^{\uparrow \downarrow}}{dx d Q^2} + \frac{d^2 \sigma^{\uparrow \uparrow}}{dx dQ^2}\right]}  = \frac{g_1(x)}{F_1(x)} \ .
$$

Fig.~\ref{world} from~\cite{Aid2013} shows the world data on $xg_1$ for the proton, deuteron and neutron ($^3$He).
The gluon distribution $\Delta g$ enters the expression for
$g_1$ only at higher order in perturbation theory; however,
it drives the scaling violations (i.e. the $Q^2$-dependence)
of $g_1(x,Q^2)$. Deep inelastic measurements hence can also
give insight into gluon polarization, provided a large lever
arm in $Q^2$ is available at fixed $x$.

\begin{figure}[!htb]
\centering\includegraphics[width=0.9\textwidth,height=0.4\textheight]{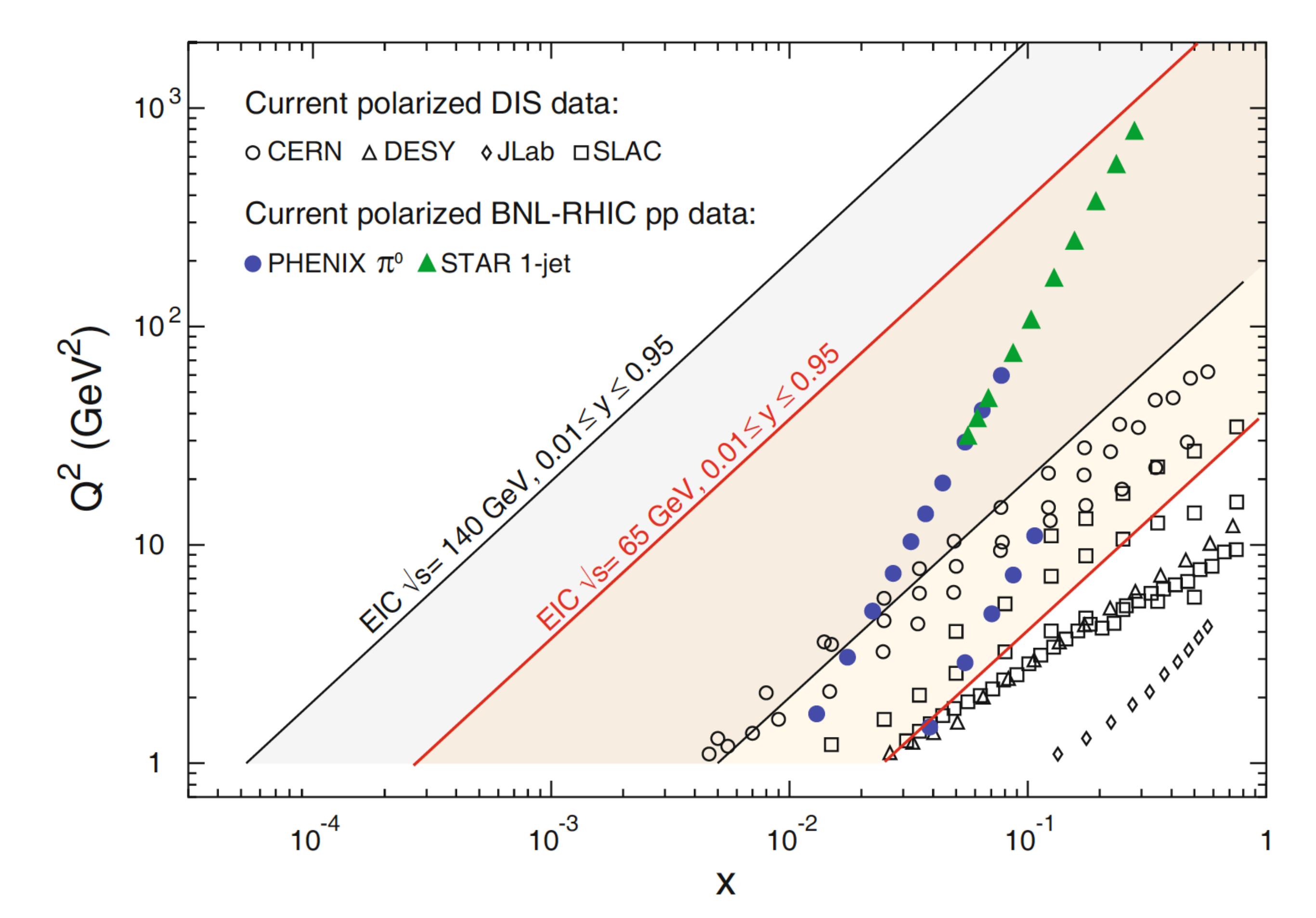}
\caption{Regions in $x$, $Q^2$ covered by previous spin experiments and anticipated to be accessible at an EIC from~\cite{Acc2016}. The values for the existing fixed-target DIS experiments are shown as data points. The RHIC data are shown at a scale $Q^2 = p^2_T$, where $p_T$ is the observed jet (pion) transverse momentum, and an $x$ value that is representative for the measurement at that scale. The $x$-ranges probed at different scales are wide and have considerable overlap. The shaded regions show the $x$, $Q^2$ reach of an EIC for center-of-mass energy $\sqrt{s}$ = 65 GeV and $\sqrt{s}$ = 140 GeV, respectively.}
\label{eickine}
\end{figure}

Fig.~\ref{eickine} provides a survey of the regions in $x$ and $Q^2$
covered by the world polarized-DIS data, which is roughly
$0.004 < x < 0.8$ for $Q^2 >$ 1 GeV $^2$. For a representative
value of $x \approx$  0.03, the $g_1(x,Q^2)$ data are in the range
$1  < Q^2 < 10$ GeV$^2$. This is to be compared to
$1 < Q^2 < $ 2000 GeV$^2$ for the unpolarized data on
$F_2(x,Q^2)$ at the same $x$. The figure also shows the vast
expansion in $x,Q^2$ reach that an EIC would provide.

\begin{figure}[!htb]
\centering\includegraphics[width=0.8\textwidth,height=0.7\textheight]{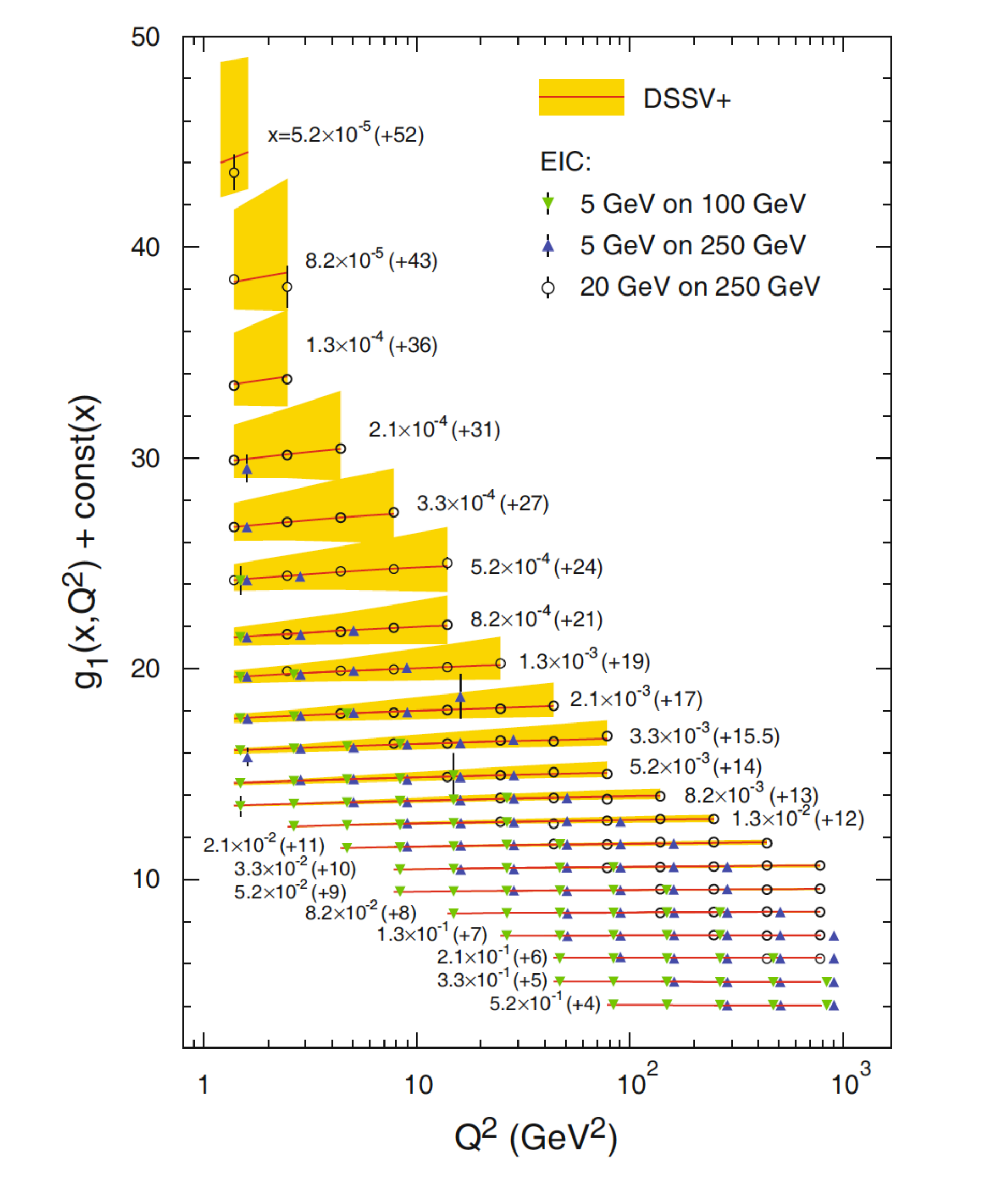}
\caption{ EIC pseudo-data~\cite{Acc2016} on the inclusive spin structure function $g^p_1(x,Q^2)$ versus $Q^2$ at fixed $x$ for 5 GeV and 20 GeV electron beams colliding with 100 GeV and 250 GeV proton beam energies at an EIC, as indicated. The error bars indicate the size of the statistical uncertainties resulting from an integrated luminosity of 10 fb$^{-1}$. The data set for each $x$ is offset by a constant $c(x)$ for better visibility. The bands indicate the current uncertainty as estimated in the ``DSSV+"
analysis.}
\label{g1p}
\end{figure}

To illustrate the power of EIC measurements of inclusive and semi-inclusive polarized deep inelastic
scattering on our knowledge of helicity parton distributions, a series of perturbative QCD analyses was performed~\cite{Asc2012} with realistic pseudo-data for various center- of-mass energies. The data simulations were based on the
PEPSI Monte Carlo generator~\cite{Man1992}. The precision of the data sets corresponds to an accumulated integrated luminosity
of 10 fb$^{-1}$ (or one to two months of running for most energies at the anticipated luminosities) and an assumed
operations efficiency of 50\%. A minimum $Q^2$ of 1 GeV$^2$ was imposed, as well as $W^2 >$ 10 GeV$^2$, a depolarization factor of the virtual photon of $D(y) > 0.1$, and $0.01 \le y \le 0.95$. Fig.~\ref{g1p} shows the pseudo-data for the inclusive structure function $g^p_1(x,Q^2)$ vs. $Q^2$ at fixed $x$.
\newpage
\section{Polarized $^{\mathbf 3}$He Ions in RHIC}

\begin{figure}
\centering\includegraphics[width=0.9\textwidth,height=0.5\textheight]{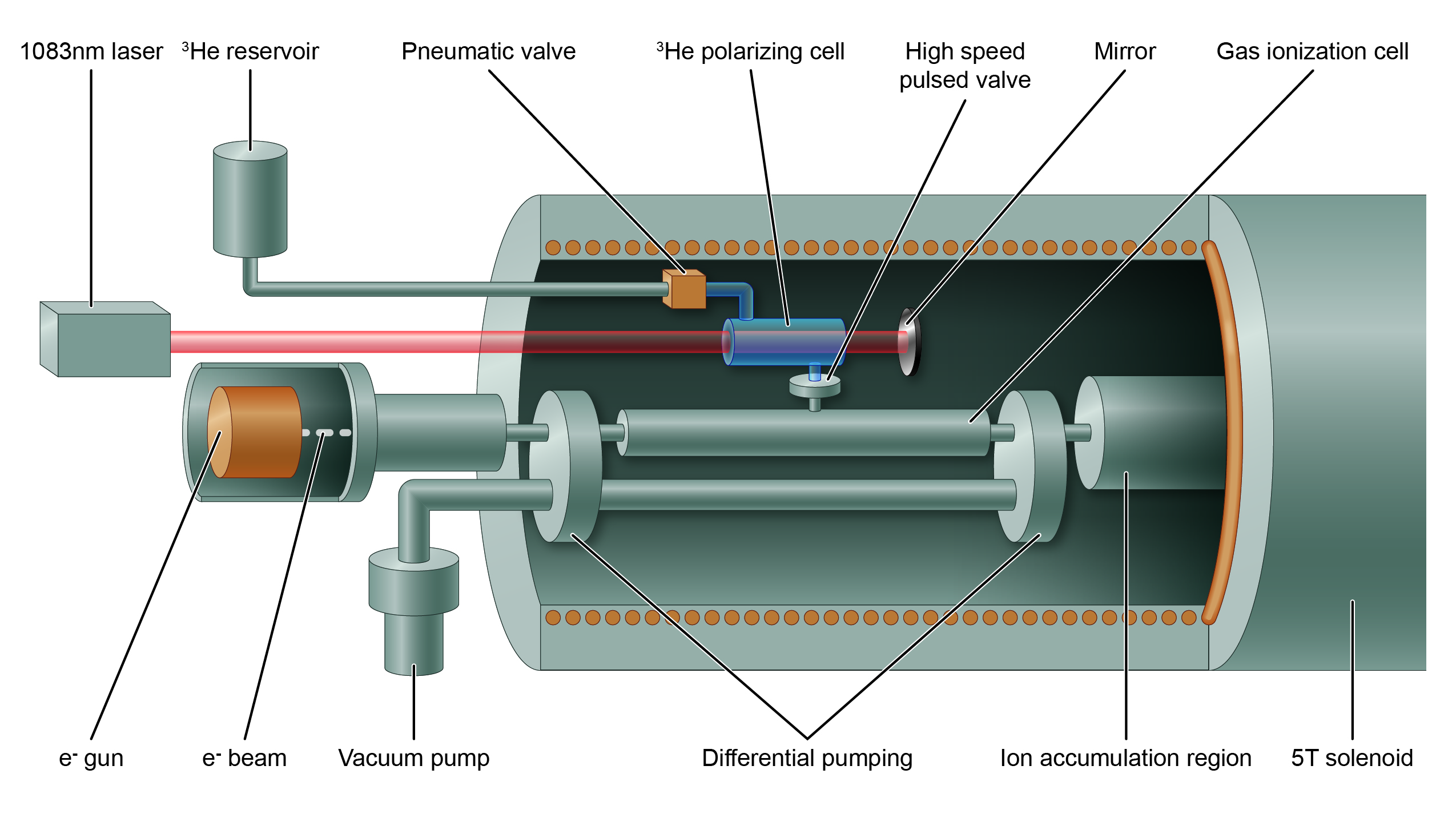}
\caption{Schematic layout of polarized $^3$He ion source under development by a BNL-MIT collaboration using optically pumped polarized $^3$He atoms directed into the existing Electron Beam Ionization Source.}
\label{source}
\end{figure}

A polarized $^3$He ion source, shown schematically in Fig.~\ref{source}, is under development by a BNL-MIT collaboration~\cite{Jon2017} motivated by the science of EIC with the goal of having polarized $^3$He beams with 70\% polarization circulating in RHIC in the early 2020s.
Following~\cite{Bai2011}, to store both protons and $^3$He ions in RHIC, the $\gamma$-factors must be matched.  The highest $\gamma$ attainable in RHIC is 177.  The maximum energy for the $^3$He ion is then 166 GeV/nucleon.  The proton$-^3$He collision luminosity $L$ is given by~\cite{Bai2011}
$$
L = \frac{N_p N_{He3} f_{rev} N_c}{2 \pi \sqrt{\sigma_{p,x}^2 + \sigma_{He3,x}^2} \sqrt{\sigma_{p,y}^2 + \sigma_{He3,y}^2}} \ .
$$
Assuming a proton bunch intensity of $2 \times 10^{11}$, a $^3$He bunch intensity of $1 \times 10^{11}$, $N_c$ =110, both round beams with a normalized emittance of 15 $\pi$ mm$-$mrad and $\beta^\star$ = 0.7 m at the IP, the peak luminosity is $1.3 \times 10^{32}$ cm$^{-2}$ s$^{-1}$.  This is about a factor of two lower than the luminosity for proton-proton collisions.

For eRHIC, the maximum energy of the $^3$He is again 166 GeV/nucleon ($\gamma$ =177).  Again, the expectation is that
the eRHIC electron$-^3$He luminosity is about a factor of two lower than the eRHIC electron$-$proton luminosity. 

Given the maximum energy for $^3$He in RHIC is 166 GeV/nucleon, the maximum center-of-mass energy and minimum $x$ attainable is less extensive than with the proton beam.  For 20 GeV electrons on 166 GeV/nucleon $^3$He in eRHIC, we have
$$
\sqrt{s}_{max} = 115 \ \ {\rm GeV}  \ \ \ {\rm and,} \ \ \ x_{min} = 7.6 \times 10^{-5} \ \ {\rm at}\  Q^2 =1 \ \ {\rm GeV}^2 \ .
$$
\begin{table}[htb]
\centering  
\caption{Comparison of $e-p$ vs. $e-^3$He luminosity, maximum center-of-mass energies and minimum $x$ in eRHIC with 20 GeV electrons.}
\vskip 0.2 true in
\begin{tabular}{|c|c|c|c|}
\hline\hline
Parameter & Unit & $e-$p & $e-^3$He \\  
      &   &    & \\
\hline
Luminosity & cm$^{-2}$ s$^{-1}$   &  $2.5 \times 10^{32}$   & $1.3 \times 10^{32}$\\ 
\hline
Max. Ion Energy & GeV  & 250 & 166 \\
\hline
$\sqrt{s_{max}}$    & GeV  &  140  & 115 \\
\hline
$x_{min}$    &   &$5 \times 10^{-5}$  & $7.6 \times 10^{-5}$ \\
\hline \hline
\end{tabular}
\label{comp}
\end{table}

\begin{table}[htb]
\centering  
\caption{Magnetic rigidity of incident $^3$He and final-state spectator $^2$H and $^1$H in DIS from polarized $^3$He at highest energy available in eRHIC with 20 GeV electrons.}
\vskip 0.2 true in
\begin{tabular}{|c|c|c|c|}
\hline\hline
Nucleus & Momentum & Charge & Magnetic Rigidity ($p/q$) \\  
              &     GeV/c     &   e        & GV/c\\
\hline
$^3$He & 498             &  $+2$   & 249\\ 
\hline
$^2$H   & 332            & $+1$    & 332 \\
\hline
$^1$H   & 166            &  $+1$  & 166\\
\hline \hline
\end{tabular}
\label{comp}
\end{table}

\subsection{Neutron Spin Structure Function $\mathbf {g^n_1(x,Q^2)}$}
\begin{figure}[!h]
\centering\includegraphics[width=0.4\textwidth,height=0.15\textheight]{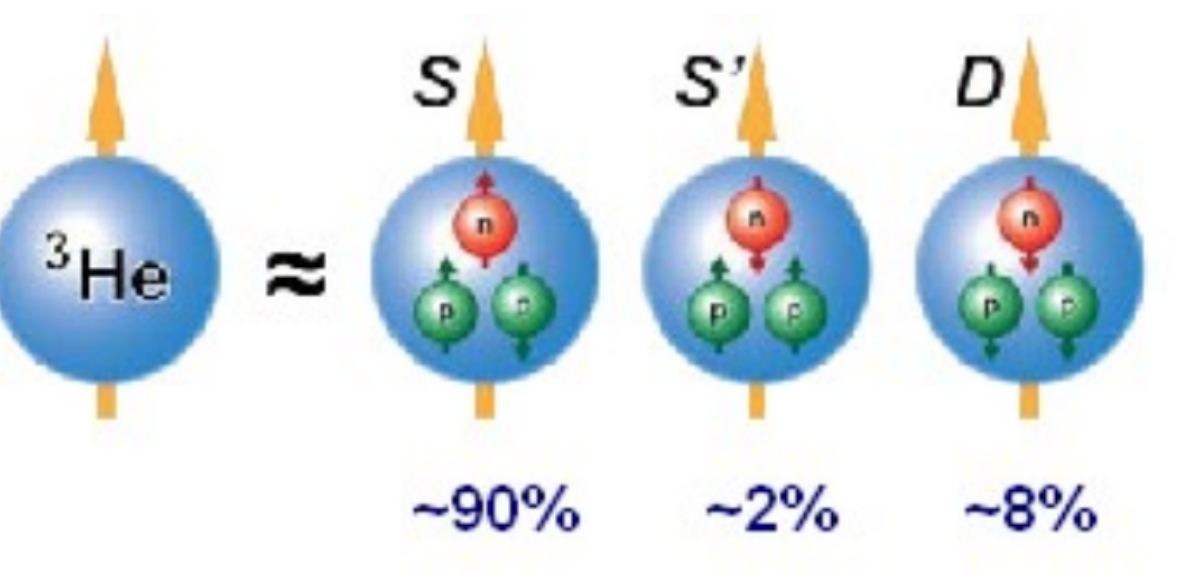}
\caption{Spin structure of the $^3$He nucleus dominated by the $S-$, $S^\prime-$ and $D-$state components of the ground state wavefunction.}
\label{he3spin}
\end{figure}

Polarized $^3$He is a well-established effective polarized neutron target for electron scattering experiments.  The ground-state nuclear spin-$\frac{1}{2}$ is well explained by the dominant $S-$, $S^\prime-$ and $D-$state wave function components illustrated schematically in Fig.~\ref{he3spin}.  For a fully polarized $^3$He nucleus, the neutron has a polarization of 87\%, while the protons have a slight residual polarization of $-2.7\%$.  Electron scattering experiments from polarized $^3$He to extract neutron structure information have been carried out at MIT-Bates, Mainz, DESY/HERMES, SLAC/E142, and Jefferson Lab.  It is clear from the previous section that inclusive spin-dependent DIS measurements on polarized $^3$He can yield very precise measurements of $g^n_1(x,Q^2)$ comparable (within factors of 2) to those expected of $g^p_1(x,Q^2)$ as described in Fig.~\ref{g1p} above.  The minimum $x$ attainable in DIS kinematics on polarized $^3$He is $7.6 \times 10^{-5}$ compared to $5 \times 10^{-5}$ on the proton.

\subsection{Detecting the Spectator Proton or Deuteron in Spin-dependent DIS from Polarized $^3$He}

\begin{figure}
\centering\includegraphics[width=0.6\textwidth,height=0.4\textheight]{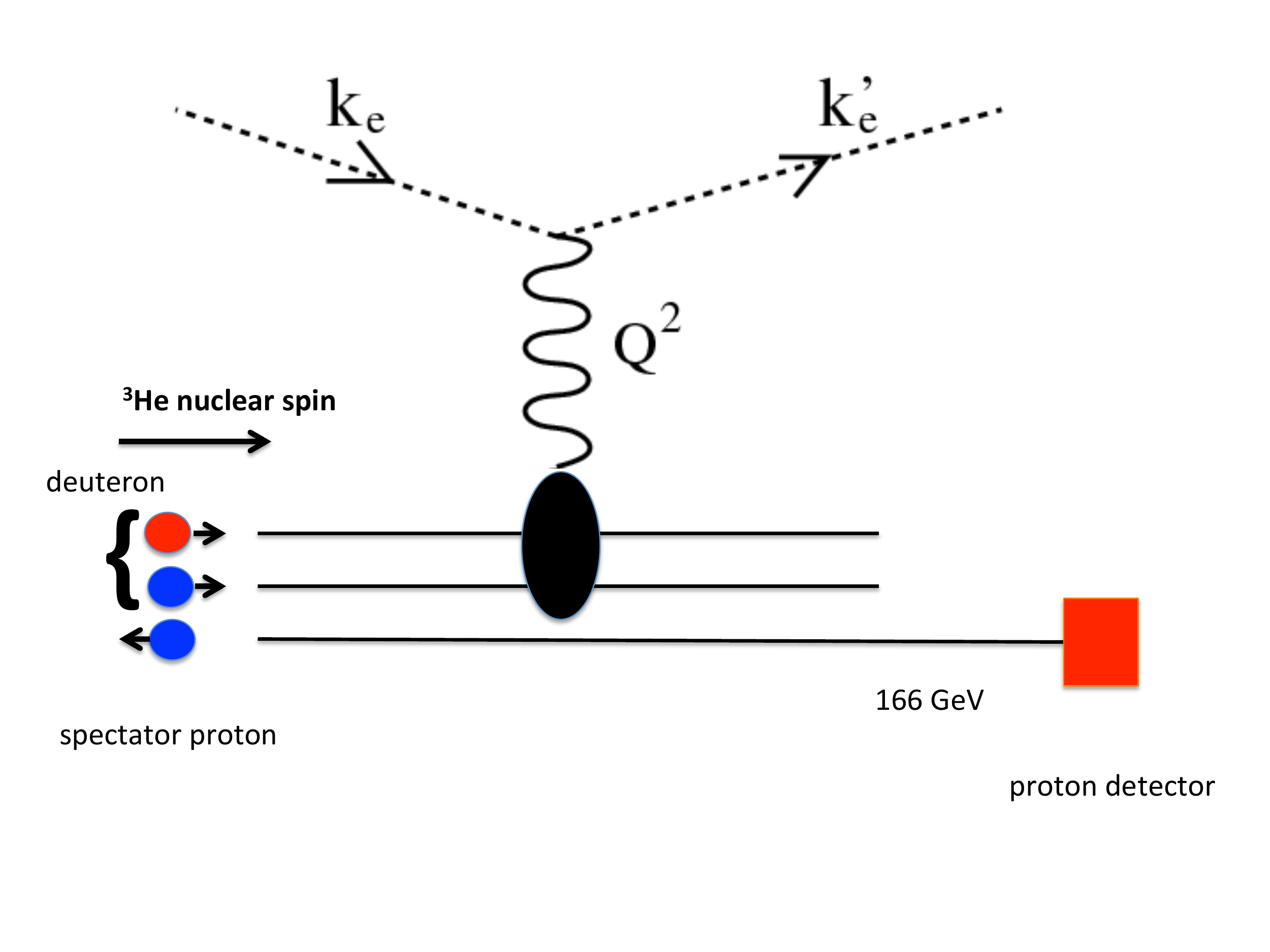}
\caption{Spin-dependent DIS from the deuteron in $^3$He at maximum energy in RHIC by tagging the spectator proton.  The deuteron is predominantly polarized in the direction of the $^3$He nuclear spin.}
\label{tagdeut}
\end{figure}

It is interesting to consider what additional observables might be measurable with detection of the spectator proton and/or deuteron in spin
-dependent electron scattering from polarized $^3$He at EIC.  The possibility to access electron-deuteron scattering by detecting (``tagging") the spectator proton is shown schematically and naively in Fig.~\ref{tagdeut}.  

To understand corrections to this simplistic picture, we consider how the spin of the polarized $^3$He nucleus arises when it is in the simplest $S-$state ($P= 88.6\%$):
\begin{eqnarray}
|^3{\rm He} \uparrow> &=& (n\uparrow) \left[ (p\uparrow p \downarrow) - (p \downarrow p \uparrow) \right] \\
&=& (n\uparrow p\uparrow)_{(J=1, M=1)} (p\downarrow) - (n\uparrow p\downarrow)_{(J=0\ {\rm or}\ 1, M=0)} (p\uparrow) \ . 
\end{eqnarray}
For the $np$ system, we have $J=1,0$ with 
\begin{eqnarray}
|1,1> &=& (n \uparrow p \uparrow) \\
|1,0> &=& \frac{1}{\sqrt{2}} \left[ (n \uparrow p \downarrow + n \downarrow p \uparrow \right] \\ 
|1,-1> &=& (n \downarrow p \downarrow) \\ 
|0,0> &=& \frac{1}{\sqrt{2}} \left[ (n \uparrow p \downarrow - n \downarrow p \uparrow) \right]  \ . \\ 
\end{eqnarray}
We can then write
\begin{eqnarray}
(n \uparrow p \downarrow)_{(J=1,M=0)} &=& \frac{1}{\sqrt{2}} \left[ |1,0> + |0,0> \right] \\
(n \downarrow p \uparrow)_{(J=0,M=0)} &=& \frac{1}{\sqrt{2}} \left[ |1,0> - |0,0> \right] \ , 
\end{eqnarray}
which allows us to express the $^3$He $\uparrow$ spin- state as
$$
|^3{\rm He} \uparrow> = |1,1> (p\downarrow) - \frac{1}{\sqrt{2}} \left[ |1,0> + |0,0> \right] (p \uparrow) \ .
$$
When normalized, this becomes
$$
|^3{\rm He} \uparrow> = \frac{1}{\sqrt{2}} |1,1> (p\downarrow) - \frac{1}{2} \left[ |1,0> + |0,0> \right] (p \uparrow) \ .
$$
Similarly, it follows that
$$
|^3{\rm He} \downarrow> = \frac{1}{\sqrt{2}}|1,-1> (p\uparrow) - \frac{1}{2} \left[ |1,0> - |0,0> \right] (p \downarrow) \ .
$$
\begin{figure}[!h]
\centering\includegraphics[width=0.8\textwidth,height=0.5\textheight]{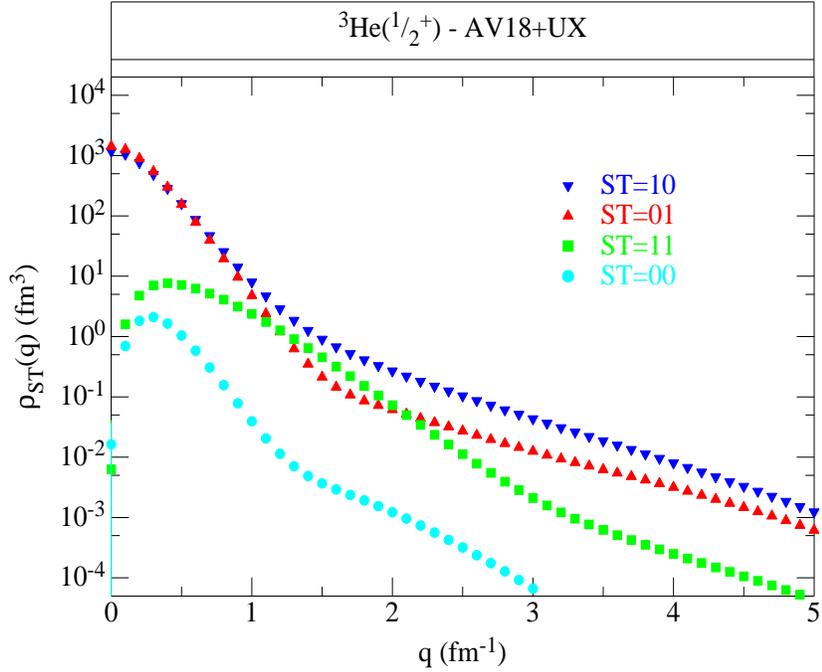}
\caption{Momentum distributions of nucleon-nucleon pairs by spin ($S$) and isospin ($T$) in $^3$He in fm$^3$ calculated using variational Monte-Carlo techniques from~\cite{Wir2014}.}
\label{he3_qST}
\end{figure}

Fig.~\ref{he3_qST} from~\cite{Wir2014} shows the momentum distributions of the different spin-isospin combinations for nucleon-nucleon pairs in $^3$He.   The number of deuterons $N_d$ in $^3$He is found to be 1.38. The number operator for the isospin $T = 0$ pairs is
$$
\frac{1}{4} \sum_{i <j} (1 - {\boldsymbol \tau}_i \cdot {\boldsymbol \tau}_j) \ ,
$$
and so $^3$He contains two $np$ pairs (1.5 $T=0$  and 0.5 $T=1$) as well as one $pp$ pair ($T=1$).  It appears that about 92\% of the $T=0$ pairs in $^3$He are in the deuteron state.  These calculations also imply that the $d + p$ components account for $\sim 70 \%$ of the proton momentum distribution in $^3$He. At small $k$ the $N_{dp}(k)$ is $\sim$80\% of the $N_p(k)$.

Recall that in terms of the states $^{2S+1}$L$_J$ of the $NN$ system, the $ST$ states can be written as 
\begin{eqnarray}
ST &=& 10 \ \ \ {\rm is} \ \ \ ^3S_1-^3D_1 \\
ST &=& 01 \ \ \ {\rm is} \ \ \ ^1S_0 \\
ST &=& 11 \ \ \ {\rm is} \ \ \ \frac{1}{15}(3 \times {}^3P_0 + 5 \times {}^3P_1 + 7 \times {}^3P_2 ) \\
ST &=& 00 \ \ \ {\rm is} \ \ \ ^1P_1 \ . \\
\end{eqnarray}
Referring to Fig.~\ref{he3_qST}, we have for $np$ pairs in terms of the $\rho_{ST}$ 
\begin{eqnarray}
\rho^{n,p}_{J=1}(q) &=& \rho_{10}(q) + \rho_{00}(q) + \frac{1}{9} \times  \rho_{11}(q) \\
\rho^{n,p}_{J=0}(q) &=& \rho_{01}(q) + \frac{1}{5} \times  \rho_{11}(q) \ . \\
\end{eqnarray}

\begin{figure}[!h]
\centering\includegraphics[width=0.9\textwidth,height=0.3\textheight]{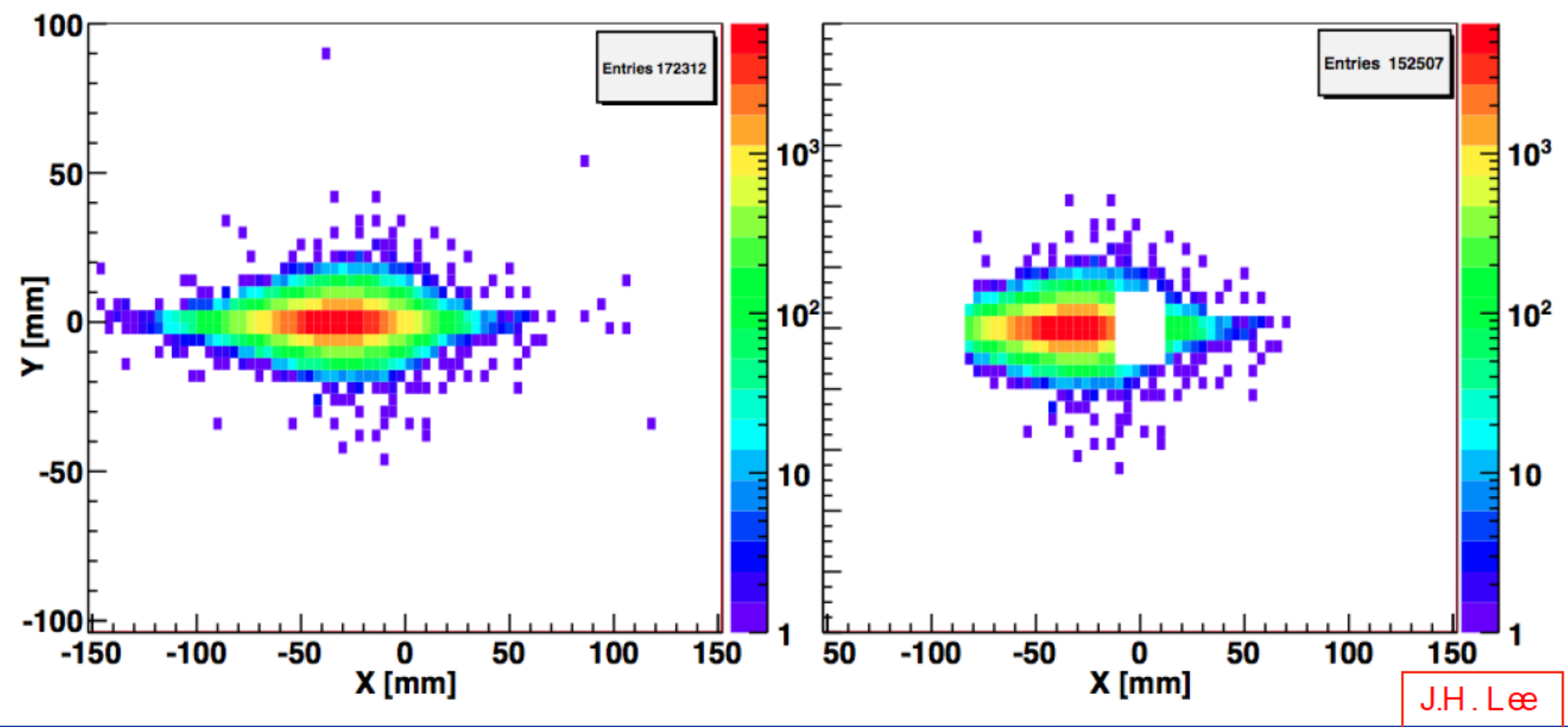}
\caption{Acceptance at the Roman Pot detector located 20 m from the IP for spectator protons in 5 GeV electrons incident on 100 GeV/nucleon $^3$He as determined in a simulation for eRHIC using DPMJET III by J.H. Lee (BNL) as reported in~\cite{Bue2014}.}
\label{pots}
\end{figure}

Spin-dependent DIS from polarized $^3$He is an incoherent sum of contributions from the four $np$ states.  
Consider now spin-dependent DIS where the spectator proton or deuteron is tagged in the final state.
 \begin{itemize}
\item{\bf Tagged deuteron:}  Scattering from the $|0,0>$ state cannot contribute. Thus,  measurement of $\overrightarrow{^3\rm He}({\overrightarrow e},e^\prime d_{\rm spectator})$ in DIS kinematics is equivalent to scattering from a negatively polarized proton 66\% of the time and 33\% of the time from a positively polarized proton.  This is equivalent to scattering from the polarized proton in $^3$He with $-$33\% polarization. This makes polarized $^3$He an effective polarized proton target.
\item{\bf Tagged proton:} 50\% of the time, the scattering arises from the $|1,1>$ state, 25\% from the $|1,0>$ state and 25\% from the $|0,0>$ state. In forming the spin-asymmetry $A$ in the DIS process $\overrightarrow{^3\rm He}({\overrightarrow e},e^\prime p_{\rm spectator})$
there will be a contribution from scattering from the deuteron $A_{ed}$, the contribution arising from the $|1,0>$ state will cancel  and there will a correction arising from a contribution $A_{corr}$ from scattering from the $np$ pair in the $|0,0>$ state, i.e.  
\begin{eqnarray}
A \sim \frac{2}{3}A_{ed} + \frac{1}{3} A_{corr} \ .
\end{eqnarray}
How large is $A_{corr}$?
\end{itemize}
Measurements have been carried out at the STAR detector at RHIC where the forward protons have been tagged~\cite{Lee2009}.
Spectator protons follow the magnet lattice and the minimum angle is determined by the beam width.  Roman Pot detectors consisting of $200 \times 100$ mm$^2$ sensitive silicon detector area located 15 mm from the beam have been considered.   Simulations have been carried out for eRHIC of 5 GeV electrons incident on 100 GeV/nucleon $^3$He ions~\cite{Bue2014}.  
Fig.~\ref{pots} shows the momentum distribution of spectator protons from $^3$He measured in a Roman Pot setup at 20 m from the interaction pint.  The acceptance for spectator protons is determined to be about 90\%.

\begin{figure}[!h]
\centering\includegraphics[width=0.6\textwidth,height=0.6\textheight,angle=-90]{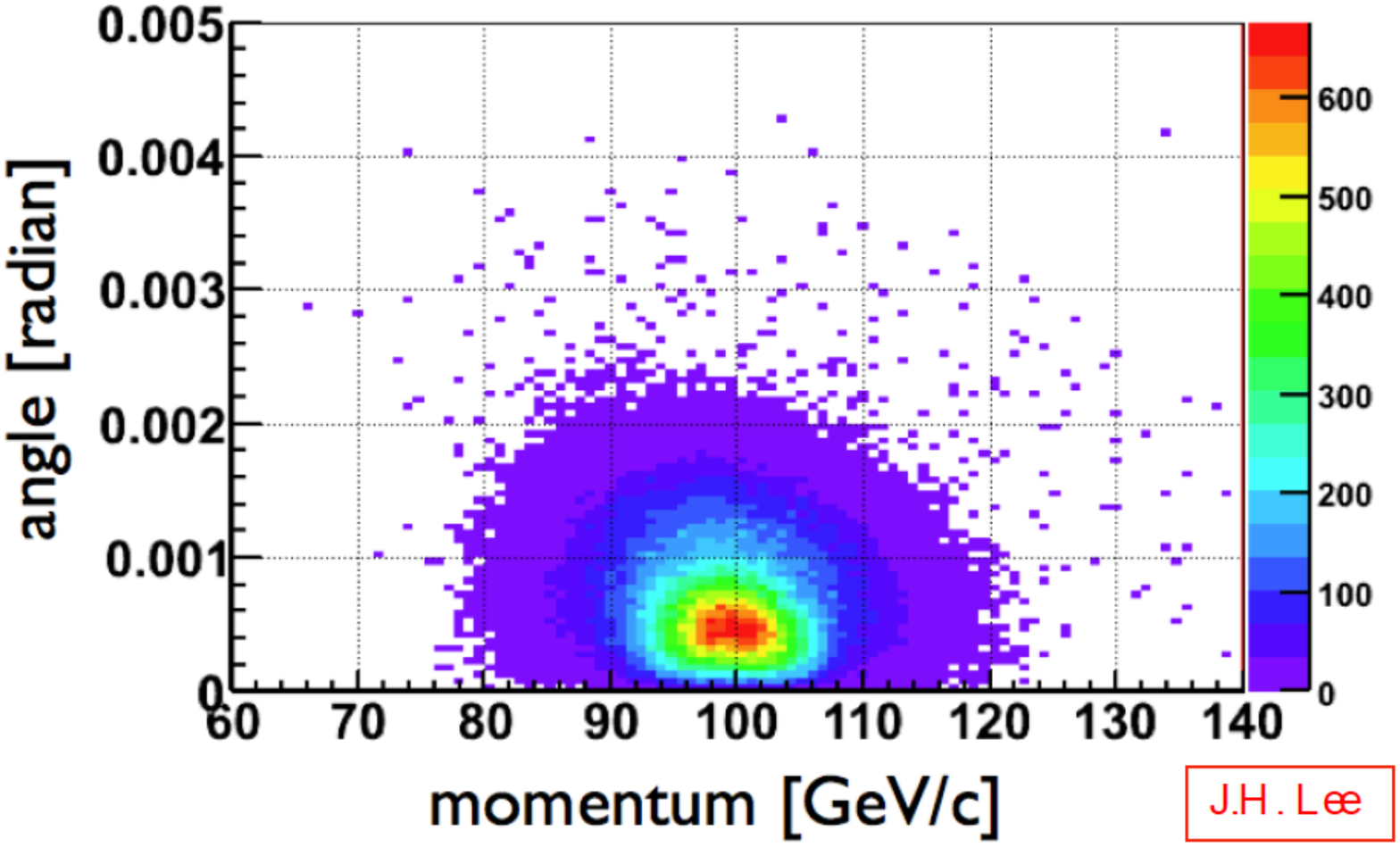}
\caption{Momentum distribution of spectator protons in 5 GeV electrons incident on 100 GeV/nucleon $^3$He as determined in a simulation for eRHIC using DPMJET III by J.H. Lee (BNL) as reported in~\cite{Bue2014}.}
\label{pots1}
\end{figure}
\section{Path Forward}
Clearly, inclusive spin-dependent DIS from polarized $^3$He will give a precision determination of $g^n_1(x,Q^2)$ over a large kinematic region that can be used with the measurements on the proton to test the Bjorken Sum Rule.   Precision high-energy $^3$He polarimetry will be essential for this as well as installation of the equipment necessary to manipulate and maintain the spin at high energies..

In considering the tagged measurements, it has struck the author that spin-dependent electron-$^3$He (or $^2$H) DIS has not been studied theoretically in the laboratory frame of EIC.  Certainly, all experiments to date have taken place in the frame where the polarized $^3$He (or $^2$H) nucleus is at rest.  The calculations to date of the spin structure of few-body nuclei are all done in the rest frame of the nucleus.   Before one can form any conclusions about the usefulness of tagging deuterons and protons, it will be necessary to carry out a calculation in the EIC laboratory frame.  This will involve  boosting from the nuclear rest frame to the collider frame. It should aim to answer some basic questions: 
\begin{itemize}
\item{} What is the probability of detecting a spectator deuteron or proton when the partner proton or deuteron, respectively, suffers a DIS scattering from a polarized high-energy electron?
\item{} How large are the corrections in extracting $A^p_1(x,Q^2)$ and $A^d_1(x,Q^2)$ from electron scattering from polarized $^3$He?  Is it even tractable?
\item{} In spin-dependent scattering from the polarized few body nuclei, e.g. $^6$Li, can one extract $A^d_1(x,Q^2)$ by tagging on the spectator alpha particle?
\end{itemize}
The one calculation to date on spin-dependent electron scattering in the EIC laboratory frame has been of elastic electron-proton scattering by Sofiatti and Donnelly~\cite{Sof2011}.  They find substantial spin effects when boosting from  the rest frame of the proton to the collider frame.

\acknowledgements

I thank Wolfram Fischer for guidance on realistic RHIC running parameters, Robert Jaffe for discussion of the $^3$He spin structure, Robert Wiringa for his detailed calculations on the spin structure of the $np$ pairs, and T. William Donnelly for discussion of boost effects between the $^3$He rest frame and the EIC laboratory frame.

\end{document}